\documentclass{webofc}
\newcommand{\be}{\begin{equation}}
\newcommand{\ee}{\end{equation}}
\usepackage[varg]{txfonts}   
\begin{document}
\title{The improved Ginzburg-Landau technique}
%
%

\author{\firstname{Massimo} \lastname{Mannarelli}\inst{1}\fnsep\thanks{\email{massimo.mannarelli@lngs.infn.it}}}

\institute{INFN, Laboratori Nazionali del Gran Sasso, Via G. Acitelli,
22, I-67100 Assergi (AQ), Italy }

\abstract{%
 We discuss an innovative method for the description of inhomogeneous phases designed to improve the standard Ginzburg-Landau expansion. The method is characterized by two key ingredients. The first one is a moving average of the order parameter designed  to account for the long-wavelength modulations of the condensate. The second one is  a sum of the high frequency modes, to improve the description of the phase transition to the  restored phase. The method is applied to compare the free energies of 1D and 2D  inhomogeneous structures arising in the chirally symmetric broken phase.}
\maketitle
\section{Introduction}
\label{intro}
In this contribution we discuss the  method presented in~\cite{Carignano:2017meb} for the description of the inhomogeneous phases. The emphasis is on cold quark matter and, more in detail, on the properties of the chiral symmetry broken ($\chi$SB) phase featuring an inhomogeneous  condensate. 

\begin{figure}[h]
\centering
\includegraphics[width=10cm,clip]{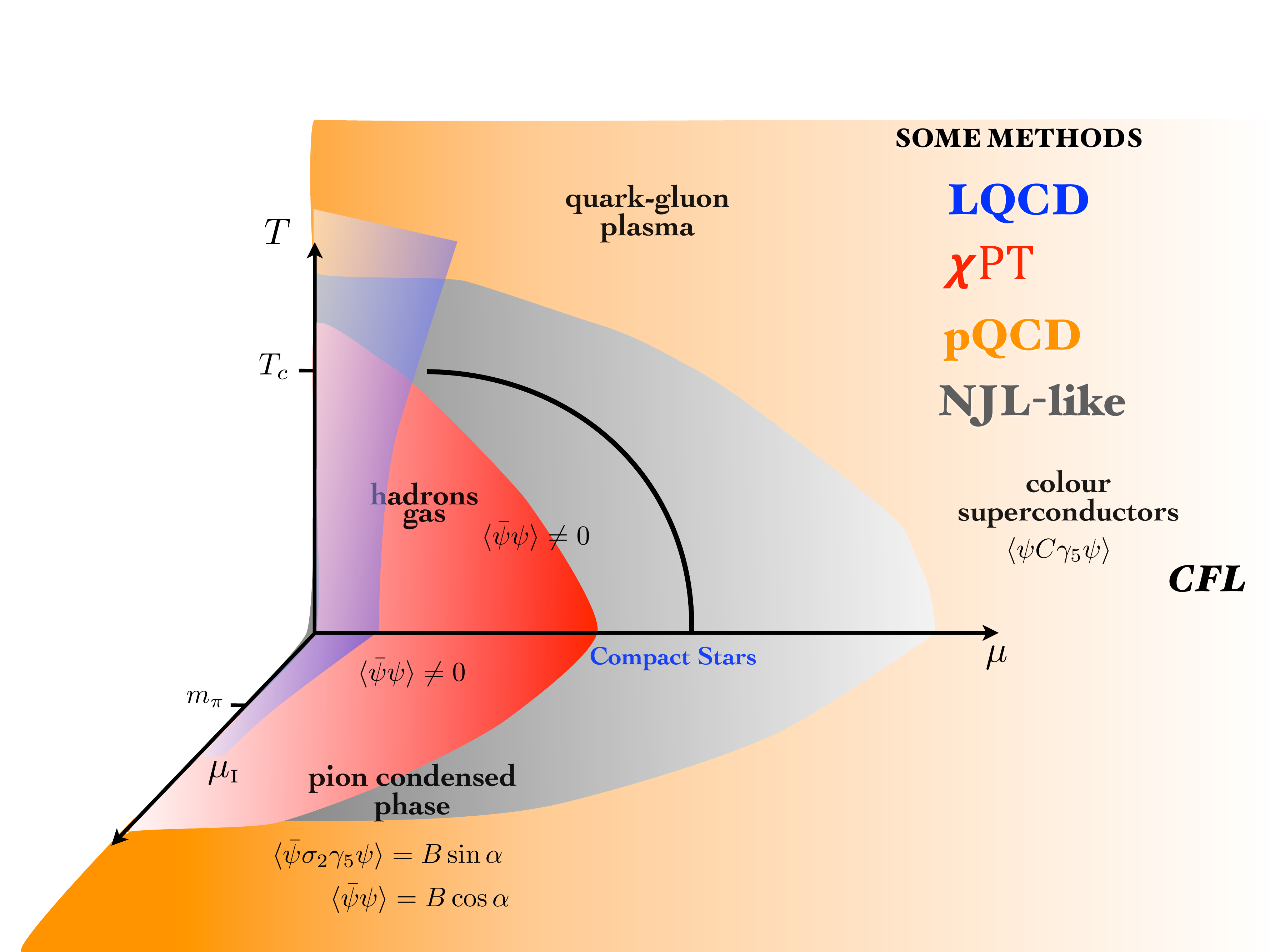}
\caption{Cartoon of the so-called QCD phase diagram:  a grand-canonical description of the properties of hadronic matter as a function of the average quark chemical potential, $\mu$, of the isospin chemical potential, $\mu_I$,  and of the temperature, $T$. The quark condensates for every phases are indicated.  The solid line corresponds to a possible first order phase transition, which is currently investigated by relativistic  heavy-ion colliders, between the confined and the deconfined phases. Shaded areas correspond to the regions of applicability of the corresponding theoretical method.   }
\label{fig:phase_diagram}       
\end{figure}

The Lagrangian describing Quantum Chromodynamics (QCD) for  three flavor massless quarks  has the following symmetries
\be
S\!U(3)_\text{c} \times \underbrace{S\!U(3)_\text{L} \times S\!U(3)_\text{R}}_{\displaystyle\supset[U(1)_\text{e.m.}]} \times U(1)_\text{B}\,, 
\ee
where $S\!U(3)_\text{c}$ is the gauge color symmetry and the other symmetries are global, apart for the $U(1)_\text{e.m.}$ subgroup of the chiral rotations describing the electromagnetic interactions. 
The ground state may have a lower symmetry because the QCD interaction leads to the formation  of various condensates occurring with  a specific  symmetry breaking pattern. Schematically, the three quark condensates that are the most relevant ones for phenomenology are the followings:
\begin{align}
\langle \bar \psi \psi \rangle & \qquad \text{Chiral condensate: Locks the chiral symmetries}  \nonumber\\
\langle \bar \psi\sigma_2 \gamma_5 \psi \rangle & \qquad  \text{Pion condensate: Locks chiral rotations and breaks $U(1)_\text{e.m.}$} \nonumber \\
\langle \psi C \gamma_5 \psi\rangle & \qquad \text{Diquark condensate: Breaks the $S\!U(3)_\text{c}$ gauge group}\nonumber\,,
\end{align}
where $\psi$ is the quark field, with spinorial, color and flavor indices suppressed.
The  various condensates should be realized in different regions of the so-called QCD phase diagram, see the cartoon in Fig.~\ref{fig:phase_diagram}, corresponding to different physical situations. For every phase  we have indicated the possible condensates and the corresponding phases. The region at high temperature  where no quark condensate is expected corresponds to the so-called quark-gluon plasma and  is the one relevant for the Quark epoch of the early Universe, corresponding to a time $t\sim 10^{-12} s$ after the big bang. This is as well the region studied by relativistic heavy-ion colliders.  Instead, at low-temperature and high quark chemical potentials the diquark condensate should be realized, see~\cite{Rajagopal:2000,Alford:2008} for  reviews,   and at asymptotic densities the color flavor locked (CFL in the figure) phase~\cite{Alford:1998mk} should be favored. The region where the chiral condensate and the pion condensate are favored correspond to the "hadron gas" and "pion condensed" phases, respectively.

As indicated in the figure, there are different theoretical methods available for studying the various phases. Perturbative methods (pQCD), relying on the asymptotic  freedom of QCD, can only be applied at very large energy scales. Chiral perturbation ($\chi$PT) theory is instead valid for low energy scales. Whenever doable, lattice simulations (LQCD) are an extremely useful tool, especially for the description of the static properties of matter.  These methods are grounded on QCD and have a well defined range of validity. When these methods fails  the alternative is to study hadronic matter using  the Nambu-Jona-Lasinio (NJL) model~\cite{Klevansky:1992qe} or similar, based on a contact interaction   having the global symmetries of QCD, but lacking of the gauge part. In this way one obtains a qualitative and semiquantitative description of the various phases, although the results depend on the parameters characterizing the model. Unfortunately, the region between the chirally broken phase and the deconfined phase at large $\mu$ can only be studied by NJL-like models.

\section{Inhomogeneous phases}
In Fig.~\ref{fig:phase_diagram} we have only indicated the homogeneous phases, but  inhomogeneous phases are expected to be favored in various regions. One important example is the  Pasta phase~\cite{Ravenhall:1983uh}  in nuclear matter at high densities. Regarding cold quark matter, inhomogeneous phases are expected
close to the phase transition between the chirally broken/chirally restored  phases and in the deconfined phase.   Model calculations indicate that two inhomogeneous phases may arise:  the crystalline color superconducting phase see~\cite{Anglani:2013gfu} for a review,  and  the inhomogeneous $\chi$SB phase  see~\cite{Buballa:2014tba} for a review. The former is realized in the deconfined phase for  neutral and beta equilibrated matter, while the latter  is  expected to arise between the homogeneous $\chi$SB phase and the chirally restored phase if the quark-antiquark coupling strength is sufficiently large. Unfortunately, both phases are expected to be in the gray-region of Fig.~\ref{fig:phase_diagram}, where no first-principle methods can be applied and one has to rely on NJL-like models.
Even within the NJL framework, studying the inhomogeneous phases is an extremely challenging task, because one needs to perform a  diagonalization of the full quark Hamiltonian, which is a complicated numerical problem. Alternatively, one can perform a Ginzburg-Landau (GL) expansion in the order parameters and its derivatives. Hereafter we focus on the inhomogeneous $\chi$SB phase, with the standard GL expansion~\cite{Nickel:2009ke,Abuki:2011pf}   given by
\begin{align}
\label{eq:Omega_GL}
\Omega_\text{GL} = & \Omega[0] +\int \frac{d{\bf x}}{V}\left[   \alpha_2 M^2 + \alpha_4 \left( M^4 + (\nabla M)^2 \right)  + \alpha_6 \left(M^6 + 3(\nabla M)^2 M^2 + \frac{1}{2} ( \nabla M^2 )^2 + 
\frac{1}{2} (\nabla^2 M)^2 \right) \right. \nonumber \\  &\left. +  \alpha_8 \left(14 M^4 (\nabla M)^2 - \frac{1}{5} (\nabla M)^4 + \frac{18}{5}M (\nabla^2 M) (\nabla M)^2 + \frac{14}{5} M^2 (\nabla^2 M)^2 + \frac{1}{5}  (\nabla^3 M)^2 \right) 
+ \dots \right]\,,
\end{align}
where $V$ is the total volume and  the various coefficients are  independent of the particular modulation considered. Sometimes these  are called "universal" coefficients because if they were  derived by the appropriate microscopic high energy theory, in the present case from QCD, they would only depend on the grand-canonical variables. Since we consider the system at vanishing temperature, they should only depend  on $\mu$. However, the $\chi$SB inhomogeneous phase is realized in the gray region of Fig.~\ref{fig:phase_diagram}, meaning that we are forced  to use  the NJL model. In this case the coefficients not only depend on $\mu$, but also on the regularization scheme.  We use  the Pauli-Villars regularization scheme~\cite{Klevansky:1992qe, Nickel:2009wj}, which allows to keep into account the high energy contributions. In this way  one obtains 
 \begin{align}
 \label{eq:alpha_n}
 \alpha_2 & = \frac{1}{4G}- \frac{N_f N_c}{8\pi^2}  \left(3 \Lambda^2 \log\left(\frac{4}{3}\right) - 2 \mu^2\right) \,, 
& \alpha_4  = -\frac{N_f N_c }{16\pi^2} \log\left(\frac{32 \mu^2}{3 \Lambda^2}\right) \,, \nonumber\\
 \alpha_6 & =\frac{N_f N_c }{96\pi^2}\left(\frac{11}{3 \Lambda^2} +  \frac{1}{\mu^2} \right) \,,
 &\alpha_8  = \frac{ N_f N_c}{256 \pi^2} \left(\frac{1}{2 \mu^4}  - \frac{85}{27 \Lambda^4} \right)  \,, 
\end{align}
where $\Lambda$ is an ultraviolet regulator and  $N_f$ and $N_c$ are the numbers of quark flavors and colors, respectively. 

The GL expansion is designed to work  close to the Lifshitz point where both $M$ and $\nabla M$  are small. Indeed, one implicitly assumes that  the terms in Eq.~\eqref{eq:Omega_GL} proportional to the same $\alpha_n$ are  equally important. However, if one wants to extend the study far from the Lifshitz point this assumption is not in general true. In particular, at vanishing temperature,  approaching the second order phase transition $M$ is expected to vanish, but  $ (\nabla M)/M$ is nonzero. Improving  the GL free energy by brute force,  including higher order terms, is not easy because the calculation of the corresponding coefficients becomes increasingly hard. 

In order to extend the GL scheme away from the  Lifshitz point we propose the `improved Ginzburg-Landau'' (IGL)  expansion~\cite{Carignano:2017meb}, which for a real order parameter reads  
\begin{align}
\label{eq:Omega_IGL}
\Omega_\text{IGL}  = &   \frac{1}{V} \int d{\bf x}\;  \Bigg[ \, \Omega_\text{hom}(\overline{M(z)^2})  + \alpha_6 \left( 3 (\nabla M)^2 M^2 +
 \frac{1}{2}  (\nabla M^2)^2 \right) \nonumber \\ 
   & + \alpha_8 \left(14 M^4 (\nabla M)^2 - \frac{1}{5} (\nabla M)^4 + \frac{18}{5} M (\nabla^2 M) (\nabla M)^2 + \frac{14}{5} M^2 (\nabla^2 M)^2 \right) 
  + \sum_{n \geq 1} {\tilde\alpha}_{2n+2}  (\nabla^n M)^{2}  \Bigg] \,,
\end{align}
and differs from the standard GL expansion for the first and the last terms in the square bracket. The  $\Omega_\text{hom}(\overline{M(z)^2})$ term is the free energy for an homogeneous order parameter, evaluated  considering the moving average of the mass function
\be
\overline{M(z)^2}= \frac{1}{\lambda} \int_{z-\lambda/2}^{z+\lambda/2} M^2(\xi) d \xi\,,
\label{eq:mroll}
\ee
where $\lambda \lesssim 1/\mu$ is an ultraviolet cutoff. The $\Omega_\text{hom}(\overline{M(z)^2})$  term resums all the $M^n$ terms and by construction gives the free energy of the homogeneous phases when $\nabla M=0$. In particular if $M$ is space independent, it gives the free energy of the homogeneous   $\chi$SB  phase. Then,  if by increasing the quark chemical potential  one enters into the inhomogeneous $\chi$SB phase by a second order phase transition, it effectively resums the long-wavelength fluctuations of the order parameter, up to the scale $\lambda$. 

\begin{figure}[h]
\centering
\sidecaption
\includegraphics[width=6cm,clip]{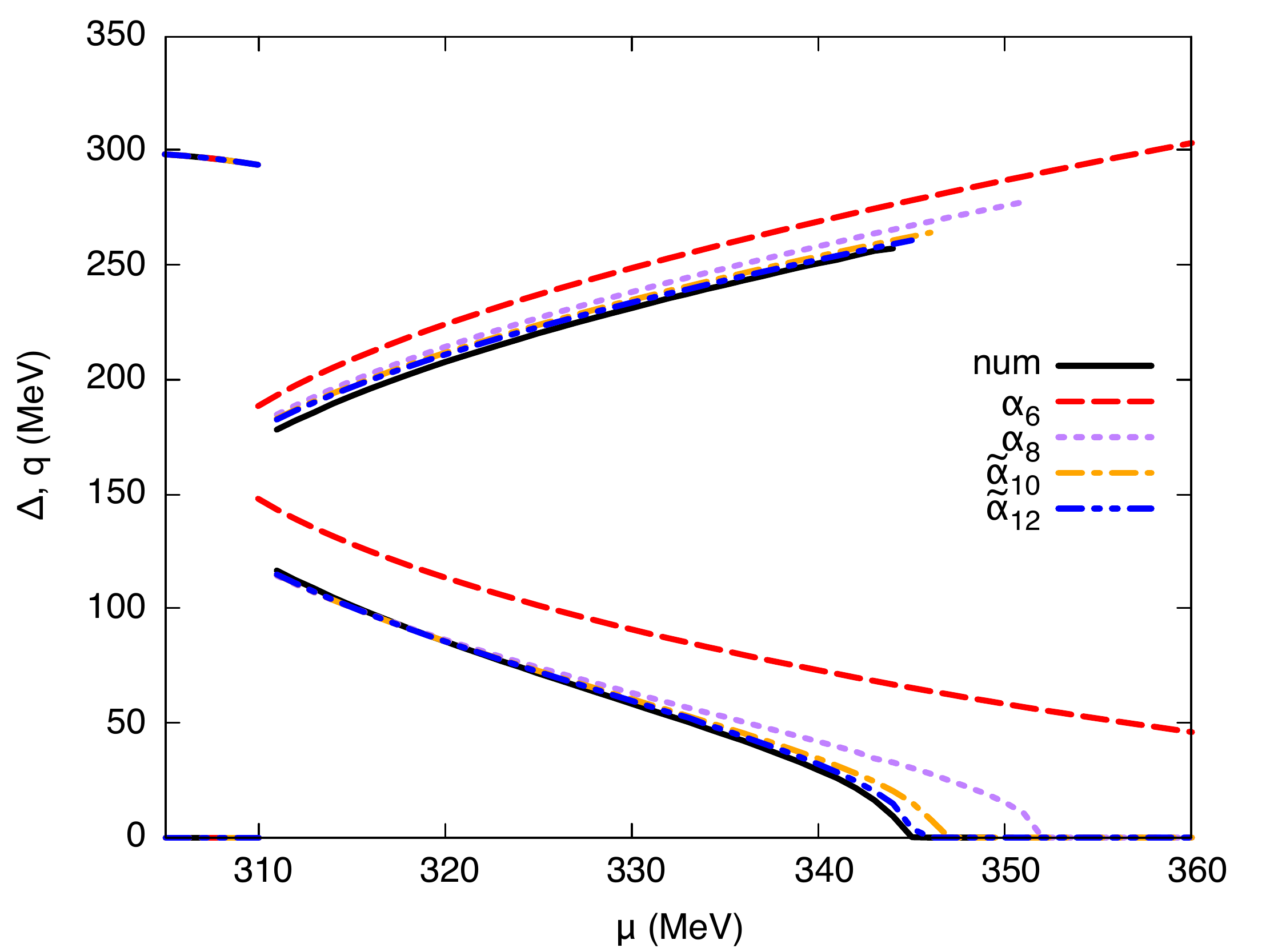}
\caption{Analysis of the IGL approximation for the plane wave ansatz in Eq.~\eqref{eq:CDW}. We report the values of $\Delta$ (curves with a decreasing values) and of $q$ (curves with an increasing values) in the ground state as a function of the quark chemical potential. The solid lines are obtained by a numerical method, the other lines correspond to the IGL expansion in Eq.~\eqref{eq:Omega_IGL}, including  gradient terms of different orders. 
With increasing  number of gradient terms the second order phase transition is increasingly well described.}
 \label{fig:TFresu2}  
\end{figure}

The second novel term of the IGL expansion, corresponding to the last term in the square bracket of Eq.~\eqref{eq:Omega_IGL}, does instead resum all the gradients of the $M^2$ terms. This term is the leading one when  $M$ is small but its derivatives are large, therefore it is dominant close to the phase transition to the chirally restored phase. The effect of the $\tilde \alpha$ terms is shown in Fig.~\ref{fig:TFresu2} for the chiral density wave (CDW) ansatz
\be
M(z) = \Delta e^{2 i q z}\,, 
\label{eq:CDW}
\ee
corresponding to a plane wave along the $z-$axis. As is clear from the figure, the inclusion of more $\tilde \alpha$ terms systematically improves the approximation close to the phase transition to the normal phase, where $q$ is large and $M$ is small. In order to compute the $\tilde \alpha$ terms one can use any modulation for which the free energy solution can be easily computed. The procedure is explained in detail in~\cite{Carignano:2017meb}.

\begin{figure*}[h]
\centering
  \includegraphics[width=6cm,clip]{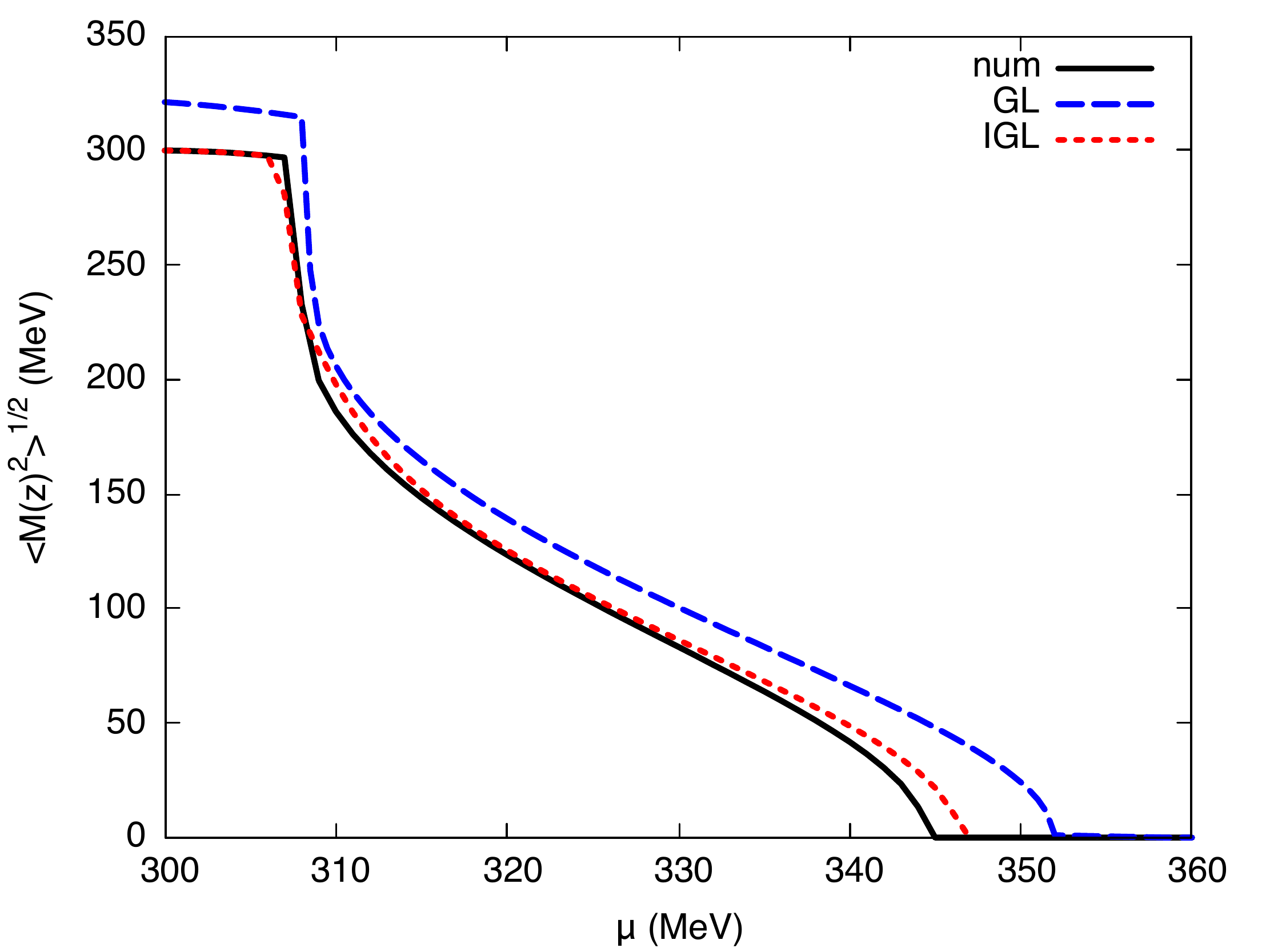}\qquad
    \includegraphics[width=6cm,clip]{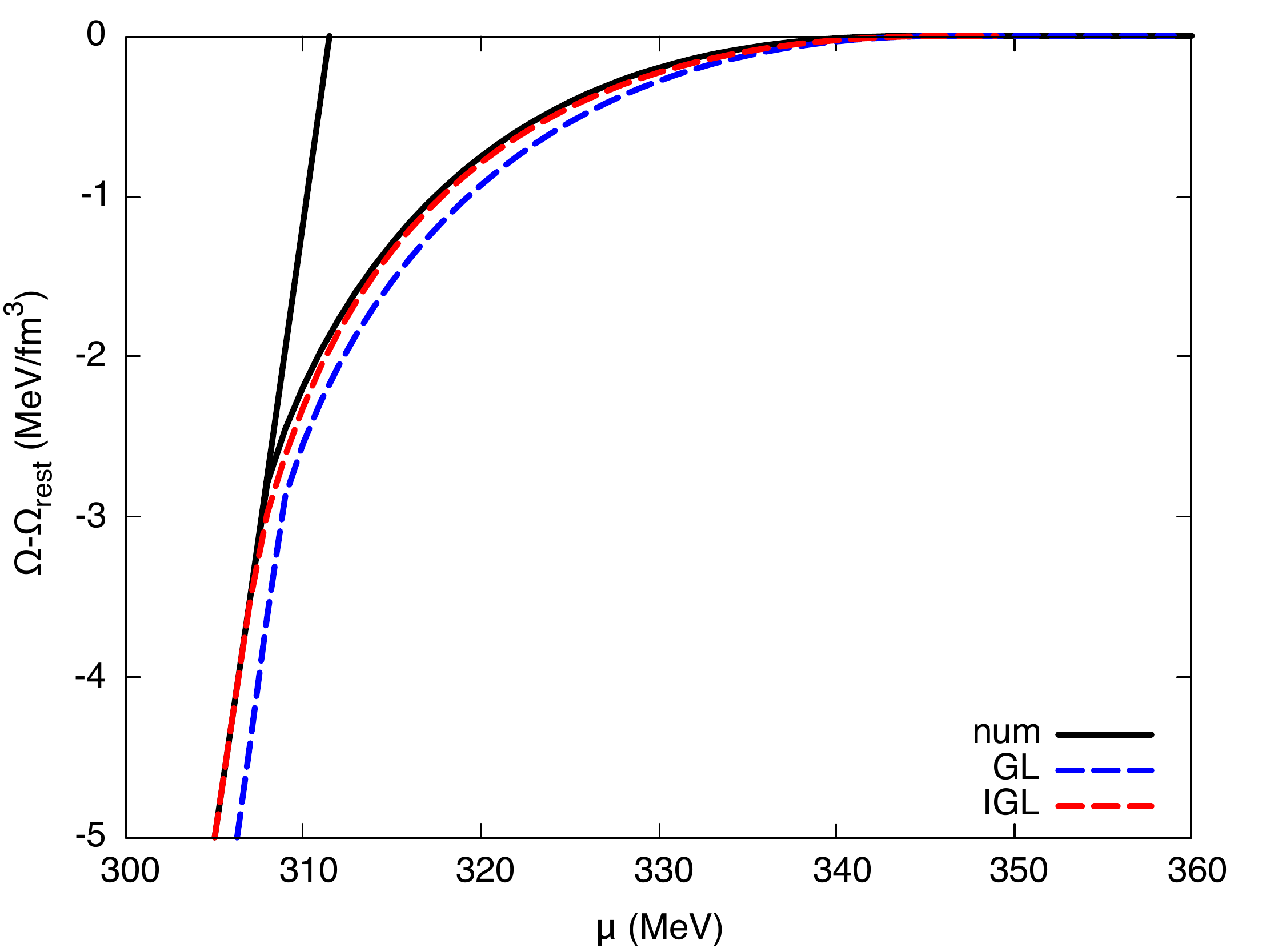}\caption{Comparison of the numerical results  with  the GL expansion and the IGL approximation  for the 1D RKC, Eq.~\eqref{eq:rkc1d}. Left:  average value of the condensate, $\sqrt{\langle M(z)^2\rangle}$. Right: difference  between the free energy at the minimum  and the free energy of the chirally restored phase. \label{fig:TFsolimprov}
}
\end{figure*}

Once the coefficients of the IGL free energy are determined, one can consider more complicated structures as the real kink crystal (RKC)~\cite{Nickel:2009wj,Buballa:2014tba}
\be
M(z) = \Delta \sqrt{\nu}\, {\text {sn}}(\Delta z | \nu)\,,
\label{eq:rkc1d}
\ee 
which is believed to be the favored spatial modulation in the $\chi$SB inhomogeneous phase. In Fig.~\ref{fig:TFsolimprov} we compare the numerical results with those obtained by the standard GL expansion and by the IGL expansion including up to $\tilde \alpha_{10}$ terms. 

As in the previous case the IGL improves the description to the normal phase, thanks to the $\tilde \alpha$ terms. Moreover, since in this case the onset of the inhomogeneous $\chi$SB phase happens by a second order phase transition, the long-wavelength term $\Omega_\text{hom}(\overline{M(z)^2})$ allows a better describe of this region.

By using the IGL method we are in a position to easily test different 2D modulations and to compare the corresponding free energies in a reliable way. We consider several structures including    a square lattice with two RKC-type modulations along the $x$ and $y$ directions, that is 
\be\label{eq:snsn}
M(x,y) = \Delta\nu {\text{sn}}(\Delta x, \nu)  {\text{sn}}(\Delta y, \nu)\,,
\ee
and the two-dimensional square lattice with a sinusoidal ansatz, that is, 
 \be\label{eq:coscos}
 M(x,y) = \Delta \cos(q x) \cos(q y) \,,
 \ee
both depending on only two parameters.

The numerical evaluation of the free energies  of these modulations is extremely complicated~\cite{Carignano:2012sx}, as it  requires an expansion of the order parameter in a large number of Fourier harmonics. Then, the minimization procedures amounts to a minimization of  the free energy with respect to all of their amplitudes. Instead, within the IGL approximation the minimization requires the evaluation of few integrals and can be straightforwardly implemented

\begin{figure}
\centering
\sidecaption
  \includegraphics[width=6cm,clip]{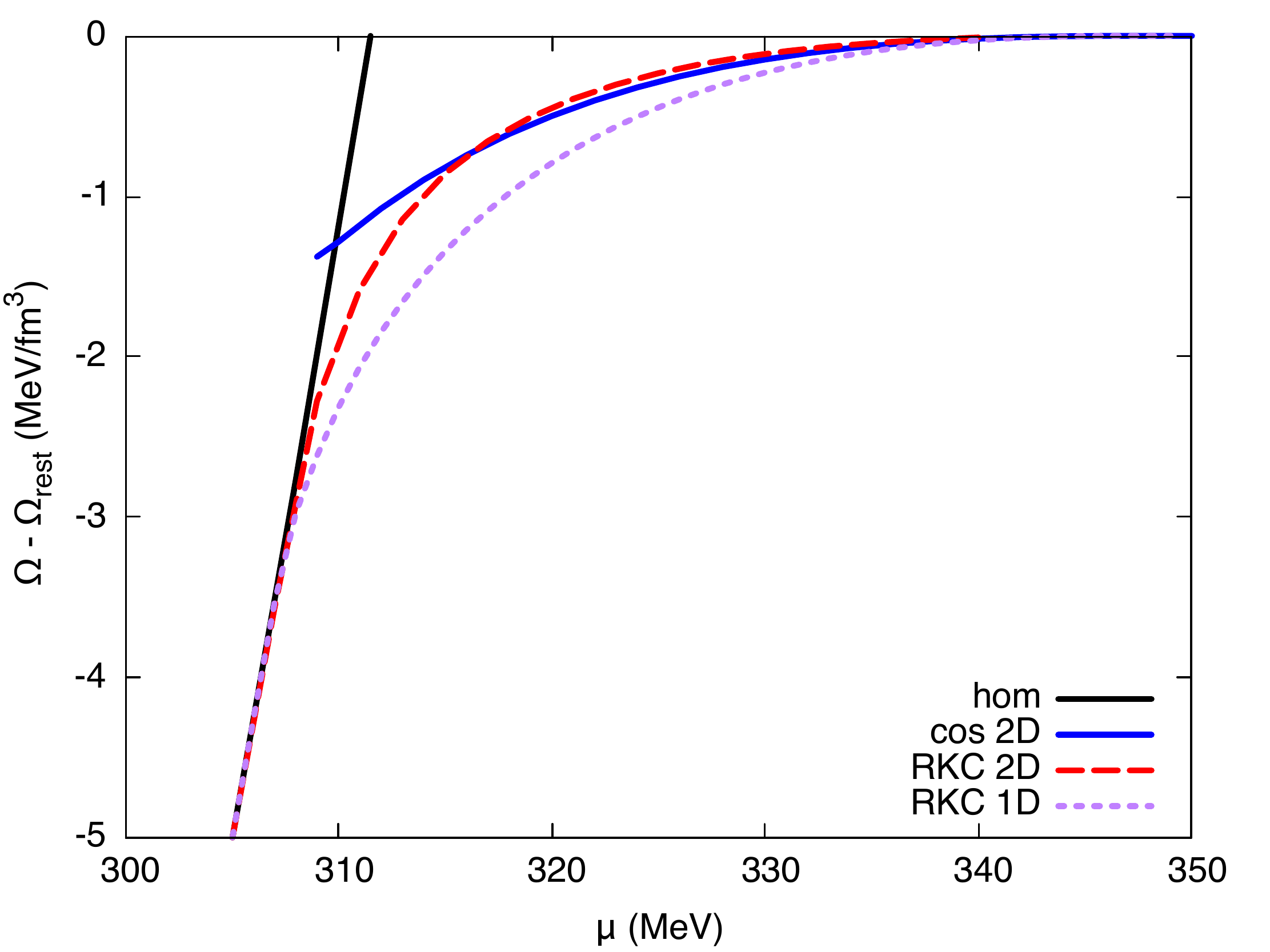}
\caption{Free energies of three different modulations as a function of the quark chemical potential. The dotted (purple) line corresponds to the 1D RKC, see Eq.~\eqref{eq:rkc1d},  the dashed (red) line indicates the 2D RKC, see Eq.~\eqref{eq:snsn}, and the solid (blue) line corresponds to the  2D cosine, see Eq.~\eqref{eq:coscos}. The 1D RKC is the energetically favored inhomogeneous phase. All the results have been obtained  within the IGL approximation in  Eq.~\eqref{eq:Omega_IGL} including up to $\tilde\alpha_{10}$ terms. 
 \label{fig:rkc2d}
}
\end{figure}

When computing the free energy associated with these two modulations we find that they are almost degenerate, apart in the region close to the onset of the inhomogeneous phase, see Fig.~\ref{fig:rkc2d}. In that figure we also see that  the 2D modulations are  disfavored with respect to the 1D RKC in Eq.~\eqref{eq:rkc1d}.  We performed a further check  by considering the ansatz 
\be
M(x,y) = \Delta \left[ \sqrt{\nu_x}\,  {\text{sn}}(\Delta x, \nu_x)  + \sqrt{\nu_y}\, {\text{sn}}(\Delta y, \nu_y) \right] \,,
\ee
which can interpolate between the 1D RKC modulation and a more involved two-dimensional structure. 
Consistent with our other results, we find that the ground state  always corresponds to one in which one of the two modulations disappears with $\nu_x$ or $\nu_y$  being zero.

\section{Conclusions}
This IGL approach can be viewed as an effective field theory based on the scale separation between long-wavelength fluctuations, dominating the transition between the homogeneous and the inhomogeneous broken  phases, and rapid fluctuations governing the transition to the  restored phase. By construction the IGL reproduces the homogeneous limits and allows for a description of the  transition to the restored phase with
arbitrarily high precision by a controlled gradient expansion.

The IGL approach has been developed in~\cite{Carignano:2017meb} and  applied to  the inhomogeneous $\chi$SB  phase at $T=0$. However, it can be extended to study other phases. For cold quark matter, the IGL method can be used to rapidly evaluate the free energy of the various crystalline color superconducting configurations considered in \cite{Rajagopal:2006ig}, eventually extending the analysis to different modulations. Clearly, one needs to obtain an expression of the IGL free energy analogous to the one in  Eq.~\eqref{eq:Omega_IGL}. This requires some work, but seems doable; in particular it should be possible to  obtain the $\tilde\alpha$ coefficients by the method described in~\cite{Carignano:2017meb} and using the simple Fulde-Ferrell  type plane wave~\cite{FF,LO,Anglani:2013gfu}.
Numerical results (for few cases) are also available~\cite{NB:2009} and can be used as a check of the method. 

A more challenging task is to modify  the IGL method to  simultaneously include the chiral and diquark condensates. This is an interesting topic, because both phases could be realized in the region of the QCD phase diagram across the solid line of Fig.~\ref{fig:phase_diagram}.  It is reasonable to expect that the crystalline color superconducting phase arises in the spatial regions  where the chiral condensate is small and the baryonic density is large. As a first step one could consider  1D chiral modulations coexisting with a cosine  color superconducting  modulation, see for example~\cite{Anglani:2013gfu}. But more complicated structures might actually be favored.

A different important issue is the role of fluctuations. One should include them to detect possible instabilities of the space modulation of the condensate~\cite{Lee:2015bva, Yoshiike:2017kbx, Hidaka:2015xza}, which seem to be particularly important for 1D modulations \cite{Landau:1969,Buballa:2014tba,Pisarski:2018bct}. The inclusion of fluctuations in the  IGL framework  could be implemented following the procedure of~\cite{Casalbuoni:2001gt,Casalbuoni:2002pa,Mannarelli:2007bs}.

\label{sec:conclusions}

\end{document}